\documentstyle[preprint,aps]{revtex}
\begin{document}
\draft
\title{Hund's Rule for Composite Fermions}
\author{J.K. Jain and X.G. Wu}
\address{Department of Physics, State University of New York
at Stony Brook, Stony Brook, New York 11794-3800}
\date{\today}
\maketitle
\begin{abstract}

We consider the ``fractional quantum Hall atom" in the vanishing Zeeman
energy limit, and investigate the validity of Hund's maximum-spin rule for
interacting electrons in various Landau levels. While it is not
valid for {\em electrons} in the lowest Landau level, there are regions
of filling factors where it predicts the ground state spin correctly
{\em provided it is applied to composite fermions}. The composite
fermion theory also reveals a ``self-similar" structure in the
filling factor range $4/3>\nu>2/3$.

\end{abstract}

%\narrowline

Interacting
electrons in two dimensions, moving under the influence of a high
transverse magnetic field, form a new state of matter, known as the
fractional quantum Hall state \cite {tsui}. The fundamental order in this
state is characterized by
the formation of a new kind of particle, called composite fermion
(CF)\cite {jain89}. A CF is an electron carrying
an even number ($2p$) of vortices; at the mean-field level, it can
also be thought of as an electron carrying $2p$ flux quanta (flux
quantum $\phi_{0}=hc/e$).  The strongly correlated liquid of
interacting electrons in the fractional quantum Hall state
is equivalent to a weakly interacing gas of CF's.  Since each CF `eats
up' $2p$ flux quanta, the CF's see a magnetic field \cite {jain89}
\begin{equation}
B^*= B \mp 2p \phi_{0} \rho \;\;,
\label{effective}
\end{equation}
where $B$ is the external field, and $\rho$ is the electron (or CF)
density per unit area. Equivalently, the CF filling factor,
$\nu^*=\phi_{0}\rho/B^*$, is related to the electron filling factor,
$\nu=\phi_{0}\rho/B$, by
\begin{equation}
\nu=\frac{\nu^*}{2p\nu^* \pm 1}\;\;.
\label{ff}
\end{equation}
Formation of CF's implies that, insofar as the low-energy dynamics is
concerned, the system behaves as though it were at a different filling
factor.  At the special electron filling factors $\nu=p'/(2pp'\pm 1)$, the
CF's fill an integer number ($p'$) of ``quasi-Landau levels", which
explains the origin of incompressibility in a partially filled Landau
level (LL), and the observation of fractional quantum hall effect (FQHE)
at precisely these sequences of filling factors. Several other
experiments find a natural explanation in
terms of CF's \cite {vjg,du}. There
is also convincing numerical evidence for the existence of
CF's. The exact low-energy spectra of interacting
electrons at $B$ look strikingly similar to those of
{\em non-interacting} particles at $B^*$ in finite system studies, and
the microscopic wave functions of the two systems are also very
closely related \cite {dev,wu}. This Rapid Communication describes another
manifestation of the existence of CF's and their effective filling factor.

A useful theoretical model for studying the FQHE numerically
was introduced by Haldane \cite {haldane}. This model consists of an
`atom', in which $N$ interacting electrons move on the surface of a
sphere under the influence of a radial magnetic field produced
by a magnetic monopole of strength $q$ at the center. The
flux through the surface of the sphere is $2q\phi_{0}$, where $2q$ is
an integer. There is also a `nucleus' of charge $+Ne$ at the center.
This atom will be called the `FQHE atom', denoted by
$(N|q)$. The Zeeman energy will be set to zero \cite {real}.
The single electron eigenstates are the monopole harmonics \cite {yang},
$Y_{q,l,m}$, where
the orbital angular momentum $l=|q|, |q|+1, ...$, and
in any given angular momentum shell, the z-component of the angular
momentum $m=-l, -l+1, ..., l$. The degeneracy of the $l$th shell is $4l+2$.
The eigenstates of the interacting many-electron system have total angular
momentum, $L$, and total spin, $S$.

Apart from the confinement of electrons to a two-dimensional surface,
the FQHE atom differs from a regular (i.e., $q=0$) atom
in that the lowest energy shell has angular momentum
$l=|q|$, and the eigenfunctions are monopole- rather than spherical-
harmonics. Nevertheless, it is tempting to speculate that
Hund's rules might prove useful for predicting the quantum
numbers of the ground state. In this work, we investigate Hund's
first rule, which will often be referred to simply as the Hund's rule,
which says that the ground state spin has the  maximum value consistent
with the Pauli exclusion principle \cite {cowan}.  It will be shown
that this rule almost  never works for {\em electrons} in the lowest shell,
but, in a range of filling factors, it predicts the ground
state spin correctly {\em when applied to composite fermions}.

For the FQHE atom $(N|q)$, the CF's see a magnetic monopole of
strength $q^*$, given by
\begin{equation}
q^* = q-N+1 \;\;.
\label{cfff}
\end{equation}
This equation is equivalent to Eq.\ (\ref{ff}) (with $p=2$)  since, in the
thermodynamic limit $N\rightarrow\infty$, $\nu=N/2|q|$
and $\nu^*=N/2|q^*|$.
Thus, the electron atom $(N|q)$ maps into the ``CF atom" $(N|q^*)$.
The microscopic wave functions for the
low-energy eigenstates of $(N|q)$ are given in terms of the low-energy
eigenfunctions of $(N|q^*)$ \cite {dev,wu}:
\begin{equation}
\chi_{q}=\Phi {\cal P} \Phi \chi_{q^*}\;\;.
\label{wfx}
\end{equation}
Here $\Phi$ is the wave function of the filled-lowest-shell state of
$(N|\frac{N-1}{2})$ {\em constructed as though electrons were
spinless}; it is completely antisymmetric with respect to exchange
of {\em any} two electrons. ${\cal P}$ projects the wave function onto
the lowest shell. $\chi_{q}$
has the same $L$ and $S$ quantum numbers as  $\chi_{q^*}$.
The mapping is known to
work for both positive and negative values of $q^*$ \cite {dev,wu}.
We emphasize that the CF mapping is not exact, but valid
only for the low-energy states of $(N|q)$ and $(N|q^*)$. Its
usefulness lies in the fact that the low-energy Hilbert space of $(N|q^*)$ is
drastically smaller than that of $(N|q)$, provided $|q^*|<|q|$,
resulting in a simplification of the problem.

Besides the CF mapping,
\begin{equation}
(N|q) \rightarrow (N|q-N+1) \;\;,
\label{cfmapping}
\end{equation}
there are two other relatively straightforward
mappings that relate a given FQHE atom to another.
The particle-hole conjugation, which will be denoted by $C$, maps
\begin{equation}
(N|q) \rightarrow (4q+2-N|q)\;\;.
\end{equation}
The other mapping, which will be denoted by $R$, relates
\begin{equation}
(N|q) \rightarrow (N|-q)\;\;.
\end{equation}
$R$ corresponds to switching the direction of the radial magnetic field,
which leaves the eigenenergies unchanged, while the eigenfunctions are simply
complex-conjugated \cite {yang}.
$C$ and $R$ are exact; they do not change either
the size of the Hilbert space or the eigenspectrum. By themselves,
they do not simplify the problem, but, as we will
see, they help the CF mapping in further reducing the Hilbert space.

In all our numerical
calculations, we will work exclusively in the Hilbert space of the topmost
partially filled shell, i.e., treat the completely filled shells as
inert, and neglect any mixing with the higher empty shells \cite {exact}.
We label the shells by $n=l-|q|=0,1,...$, and denote the number of
electrons in the $n$th shell by $N_{n}$.  For simplicity,
only even integer values of $N_{n}$ will be considered in this work.
Also, because of the exact particle-hole symmetry in the
partially filled shell, we will only consider half-filled or
less than half-filled shells, i.e., it will be assumed that
$N_{n} \leq 2(n+|q|)+1$.
All states belonging to a given $L$-$S$
multiplet are degenerate. Therefore, the following
discussion will be restricted
to the $L_{z}=S_{z}=0$ sector, with the understanding that
each state in this sector belongs to a degenerate multiplet of
$(2L+1)(2S+1)$ states.

The origin of Hund's rule in atomic physics lies in the short-range
part of the Coulomb interaction. It is most clearly
explained by modeling the Coulomb interaction by
a repulsive delta-function interaction, which is
felt by electrons only when they coincide. Let us call a state
which has zero interaction
energy for this interaction a ``hard-core" state, since
its wave function vanishes when {\em any} two electrons coincide.
All fully polarized states satisfy the hard-core property, since the
spatial part of their wave function is completely antisymmetric.
For regular $q=0$ atoms,  Hund's rule follows because  there are no other
hard-core states.  For the FQHE atom, on the other hand,
there are many other states that also satisfy
the hard-core property. This is the reason why the Hund's rule has not
been found to be useful for the FQHE atom.

The FQHE atom has been studied numerically in the past for
partially filled {\em lowest} shell \cite {rezayi,wu,zhang}, mostly in
the context of incompressible states. For the half-filled lowest-shell
atom, $(2q+1|q)$, which correspond to $\nu=1$,
the ground state is completely polarized.
This results from the fact that the only hard-core state here is the
fully-polarized state, i.e., the Hund's rule is valid. Note that the CF
theory maps the half-filled atom into itself.

As soon as we move away from the half-filled shell, Hund's rule is
{\em maximally} violated, as discovered by Rezayi \cite {rezayi}.
He found that for $(2q|q)$, which is a single
electron short of having a half-filled shell,
the quantum numbers of the low-energy states
in {\em increasing} energy are $L$-$S$ = 0-0, 1-1, ..., $q$-$q$
(also see Fig.1). That these should be the lowest energy states is
understandable since hard-core states of $(2q|q)$ occur only at these
quantum numbers, but their ordering is rather unexpected.
It is, however, explained quite  naturally by the CF theory,
which ``peels this atom off" layer by layer.  One
application of the CF mapping followed by $R$ and $C$ gives:
\begin{equation}
(2q|q) \rightarrow (2q-2|q-1)\;\;,
\end{equation}
which is an atom with two fewer electrons, but {\em again one
electron short of being half-filled}. Since the largest allowed spin
of $(2q-2|q-1)$ is $q-1$, the $q$-$q$ state is not
available for this atom. Therefore, of all hard-core states of
$(2q|q)$, the $q$-$q$ state has the highest
energy. Iteration of this process explains the
observed ordering of states. The microscopic CF wave function
for the 0-0 state of $(2q|q)$ is obtained starting
from the unique 0-0 state of $(2|1)$; the wave function for the 1-1 state
is obtained starting from the unique 1-1 state of $(3|2)$, and so on.
However, since the CF wave functions satisfy the hard-core property by
construction, they are identical to the {\em unique} hard-core
states of $(2q|q)$ at 0-0, 1-1, {\em etc.}.
For $(6|3)$, their overlaps with the exact
0-0, 1-1, and 2-2 states are 0.9991, 0.9993, and 0.9988, respectively.
The large overlaps tell us that the low-energy Coulomb states do indeed
satisfy the hard-core property to an excellent approximation.

Sondhi {\em et al.} \cite {sondhi} have shown that the
0-0 ground state of $(2q|q)$ can be interpreted
as a ``skyrmion" state. The CF theory provides a microscopic
description of this skyrmion state; in addition, it
also explains the quantum numbers of the excitations, and shows
that analogous skyrmion state does {\em not} occur near half-filled
{\em higher shells}, as one might have anticipated.

Now let us consider $(2q-1|q)$, which is {\em two} electrons short
of having a half-filled shell. Application of the CF
mapping followed by $R$ and $C$ gives:
\begin{equation}
(2q-1|q) \rightarrow (2q-5|q-2)\;\;,
\end{equation}
{\em which is again two electrons short of having a
half-filled shell}. For even $N$, iteration of
this procedure finally gives either
$(2|\frac{3}{2})$ or  $(4|\frac{5}{2})$,
which can be further simplified into
$(2|\frac{1}{2})$ and $(4|\frac{1}{2})$. In particular,
\begin{equation}
(6|\frac{7}{2})\rightarrow (2|\frac{3}{2})\rightarrow
(2|\frac{1}{2})\;.
\end{equation}
$(2|\frac{3}{2})$ has four
states 0-1, 2-1, 1-0, and 3-0. The 3-0 state does not produce any
state at $(6|\frac{7}{2})$ by Eq.\ (\ref{wfx})
since the projection of $\Phi$ times this
state onto the lowest shell is identically zero (which is a non-trivial
result, since $(6|\frac{7}{2})$ {\em does have} a hard core state at 3-0).
The atom $(2|\frac{1}{2})$ has only two states 0-1 and 1-0.
Thus, the CF theory predicts that the lowest two states of
$(6|\frac{7}{2})$ are 1-0 and 0-1, and the next state is 2-1,
in agreement with the exact spectrum of Fig.1.
The CF wave functions \cite {unique}
have overlaps of 0.9959, 0.9978, and 0.9970 with
the exact 0-1, 1-0, and 2-1 states of $(6|\frac{7}{2})$,
respectively. For $(8|\frac{9}{2})$, the ground state is
predicted to be 0-0, again in agreement with exact results \cite
{rezayi,unpublished}.

This remarkable ``self-similarity" property is exhibited by the
general atom $(N|q)$, provided the CF atom $(N|q^*)$ lies entirely
in the lowest shell, which is the case for $q\leq (3N-2)/4$.
(Using particle-hole symmetry, this corresponds to the regime
$4/3 > \nu > 2/3$ for $N\rightarrow\infty$.)  In this range,
$(2q+1-k|q)$, which is $k$ electrons short of
a half-filled shell, maps into $(2(q-k)+1-k|q-k)$,
which is also $k$ electrons short of a half-filled shell.
$(N|q)$ with $q>(3N-2)/4$ maps into an atom which involves higher
shells.

This motivates a study of interacting electrons in higher shells.
The higher shells are different from the lowest shell in one crucial
respect. In the lowest shell, it is possible to
construct {\em product} states containing a factor of $\Phi$, which
guarantees the hard-core property.  This can be done because
{\em the product of two lowest shell states} (at $q'$ and $q''$) {\em
is also
a lowest shell state} (at $q'+q''$, provided both $q'$ and $q''$ have
the same sign).  Since these product states do not
in general have maximum spin, Hund's rule is not forced by the short
range part of the interaction.
However, there is no obvious way of generalizing this construction
to higher angular momentum shells. Therefore, one may expect Hund's
rule to be valid in higher shells, which would preclude the
possibility of structure similar to that in the lowest shell.

We have numerically studied systems of electrons interacting with the
delta-function interaction in
second ($n=1$), third ($n=2$), and fourth ($n=3$) shells.
The results are qualitatively
different from that of the lowest shell. While we {\em do} find non-fully
polarized states with zero energy, there is a region around the
half-filled shell where {\em only the fully polarized states satisfy
the hard-core property}, and therefore the Hund's rule is valid.
We have empirically determined that this region is given by
\begin{equation}
max[\frac{N_{n}-1-2n}{2},0]
\leq q< q_{c} \equiv  \frac{n+2}{4}N_{n}-\frac{n}{2}\;\;.
\label{limits}
\end{equation}
For the lowest ($n=0$) shell, this equation says that the Hund's
rule is valid only for the half-filled shell, which has
been tested for up to $N_{0}\leq 10$ \cite {rezayi}.
We have tested it numerically for $N_{n}=2,$ 4, and 6
for $n\leq 6$, $n\leq 3$, and $n\leq 1$, respectively.
In {\em all} cases in higher ($n\neq 0$) shells, we have found that
{\em at $q=q_{c}$, there is only one hard-core
state which is not fully polarized}; moreover,
{\em its quantum numbers are always 0-0}. At the moment, we do not have
a detailed microscopic explanation of these results, which should
prove quite interesting. In general terms,
Eq.\ (\ref{limits}) implies that the regime of validity of the Hund's rule
grows as we go to higher and higher shells. This makes intuitive
sense, since then the presence of the monopole (of fixed strength) becomes
progressively unimportant, and the FQHE atom effectively begins to
look more and more like a regular atom. Note that Eq.\ (\ref{limits})
is always satisfied for $q=0$.

The most relevant case for the FQHE is when electrons are in the lowest shell.
Here, the Hund's rule almost never works for electrons. However,
even though Eq.\ (\ref{limits}) is
not satisfied by the {\em electron} atom $(N|q)$, it may be satisfied
by the {\em CF} atom $(N|q^*)$. Then, the Hund's rule can be applied
to the CF's to determine the ground state spin.  (Note that the
short-range part of the inter-CF interaction corresponds
to a longer range part of the inter-electron Coulomb interaction.)
There are, of course, some cases when
the CF theory determines the ground state spin completely without
using the Hund's rule. This happens when the CF's either occupy filled
shells (so that $S=0$), or there is a single CF in the partially
filled shell ($S=1/2$). In those cases, the CF theory predicts the
spin correctly \cite {wu}. Here we have numerically studied (Table I)
several electron atoms which map into CF atoms with more than one CF in the
second shell. With the exception of $(8|-1)$, all these CF atoms
satisfy Eq.\ (\ref{limits}). In these cases, application of
Hund's rule to the CF atom indeed obtains the ground state spin
correctly.  The CF atom $(8|-1)$ is at the borderline, with $|q^*|=q_c$.
Here, the Hund's rule is not required to work, but it still predicts
the ground state $S$ correctly. We construct CF wave function for
the ground state of each $(N|q)$ according to Eq.\ (\ref{wfx}) \cite
{unique}.
The energies of these wave functions, and their overlaps with the
exact Coulomb ground states of $(N|q)$ are also given in Table I.
For $(6|\frac{9}{2})$, there are three 0-1 hard-core states,
and for $(6|5)$ there are eight 1-1 hard-core states;
for other cases, we have not determined the number of hard-core
states, since the Hilbert space is too large for direct
diagonalization, but we expect that there are several hard-core states
with the same quantum numbers as the ground state. The large
overlaps thus provide strong evidence that the ground states do indeed
possess the CF structure.

How about the total angular momentum ($L$) of the ground state?
Mapping of $(N|q)$ to a CF atom already results in a substantial
reduction in the possible values the ground state $L$,
and the actual $L$ is always found to be
within this subset. Here, according to Hund's second rule,
one might expect the ground state to have the largest $L$.
We find that this is not always the case either in our calculations
or in earlier studies of spinless electrons
\cite {dev}.

In the end, we consider the implications of our work for the FQHE.
Eq.\ (\ref{limits}) implies that the
Hund's rule is obeyed in the $n$th LL ($n=0,1,...$)
in the filling factor range $2n+(2n+2)/(n+2)>\nu> 2n+2/(n+2)$.
In particular, in the second ($n=1$) LL, it is
valid for $2+2/3 < \nu < 2+4/3$, and in the third ($n=2$) LL for $4+3/2
>\nu>4+1/2$. Eq.\ (\ref{ff}) can be used to determine the regions in
the lowest LL where the Hund's rule applies to CF's.
These features might be observable in tilted-field experiments,
since the fully polarized states
are insensitive to increase in the Zeeman energy.

In conclusion, we have identified regions where Hund's maximum-spin
rule is applicable to electrons and composite fermions
in various LL's. While it is (almost) never
valid for electrons in the lowest LL, it generally predicts the ground
state spin correctly when applied to composite fermions.
This work also discovers a layered self-similar
structure in the range $4/3 > \nu > 2/3$, and underlines some
{\em qualitative} differences between the lowest and the higher LL's in
the low-Zeeman energy limit.

This work was supported in part by the Office of Naval Research under
Grant no. N00014-93-1-0880, and by the National Science Foundation under
Grant No. DMR90-20637. We acknowledge conversations with A.S.
Goldhaber, D. Jatkar, P.M. Koch, E.H. Rezayi and S.L. Sondhi, and
thank L. Belkhir for help with programming.

\begin{tabular}{|c|c|c|c|c|c|} \hline
$(N|q)$ & $(N|q^*)$& $L$-$S$ &$E_{0}$  &  $E_{CF}$  & overlap  \\ \hline
$(6|\frac{9}{2})$&$(6|-\frac{1}{2})$&0-1&-0.5423  &-0.5418&0.9956 \\ \hline
$(6|5)$&$(6|0)$&1-1 &-0.5220 &-0.5216&0.9879 \\ \hline
$(6|\frac{11}{2})$&$(6|\frac{1}{2})$&2-1&-0.5024 &-0.5017&0.9768 \\ \hline
$(8|6)$ & $(8|-1)$ &1-1  & -0.5375  & -0.5369 & 0.9807 \\ \hline
$(8|\frac{13}{2})$ &$(8|-\frac{1}{2})$ & 0-2  &-0.5230  & -0.5228  & 0.9918 \\
\hline
\end{tabular}

\vspace{5mm}

{\bf Table Caption}

Table I. This table gives the ground state quantum numbers ($L$-$S$)
and energy ($E_{0}$) for several FQHE atoms $(N|q)$.
Energy of the CF wave function ($E_{CF}$) and its overlap
with the exact Coulomb ground state are also shown.

\vspace{2cm}

{\bf Figure Caption}

Fig. 1 Low-energy spectra of $(6|3)$ and $(6|\frac{7}{2})$.
The energies are in units of $e^2\sqrt{q}/R$, where $R$
is the radius of the sphere, and include the interaction of
electrons with the positively charged nucleus. The spin of each
hard-core state is shown on the Figure.


\begin{thebibliography}{99}
\bibitem{exact} This is exact in the limit when the radius of
the sphere, $R \rightarrow 0$, because the energy difference
between successive shells is proportional to $R^{-2}$, while the
Coulomb interaction energy $\sim R^{-1}$.
\bibitem{unique} The CF states in this work do not contain any
adjustable parameters. This is because for all $L$-$S$ states at
$(N|q)$ considered here, there is only one $L$-$S$ state at
$(N|q^*)$.
\end{thebibliography}
\end{document}